\documentstyle[11pt,aaspp4]{article}

\def\la{\mathrel{\hbox{\rlap{\hbox{\lower4pt\hbox{$\sim$}}}\hbox{$<$}}}}
\def\ga{\mathrel{\hbox{\rlap{\hbox{\lower4pt\hbox{$\sim$}}}\hbox{$>$}}}}
\def\lesssim{\mathrel{\hbox{\rlap{\hbox{\lower4pt\hbox{$\sim$}}}\hbox{$<$}}}}

\slugcomment{Accepted to {\it The Astrophysical Journal}}

\begin{document}

\title{The Fading of Supernova Remnant Cassiopeia A from 38 MHz to 16.5 GHz from 1949 to 1999 with New Observations at 1405 MHz\altaffilmark{1}}

\author{Daniel E. Reichart\altaffilmark{2} and Andrew W. Stephens\altaffilmark{3}}

\altaffiltext{1}{The data presented herein were obtained by participants of Educational Research in Radio Astronomy (ERIRA) 1992 -- 1999, an outreach program that has received support from the National Radio Astronomy Observatory, the Ohio State University, the Pennsylvania State University, and the University of Pittsburgh at Bradford.}
\altaffiltext{2}{Department of Astronomy and Astrophysics, University of Chicago, 5640 South Ellis Avenue, Chicago, IL 60637; reichart@oddjob.uchicago.edu}
\altaffiltext{3}{Department of Astronomy, Ohio State University, 140 West 18th Avenue, Columbus, OH 43210} 

\begin{abstract}
We report 1405 MHz measurements of the flux density of the $\approx 320$ year old supernova remnant Cassiopeia A, relative to the flux density of Cygnus A, made between 1995 and 1999.  When compared to measurements made between 1957 and 1976, we find that the rate at which Cassiopeia A has been fading at this and nearby frequencies has changed from $\approx 0.9$ \% yr$^{-1}$ in the 1960s to $\approx 0.6 - 0.7$ \% yr$^{-1}$ now.  Furthermore, we have collected from the literature measurements of this fading rate at lower (38 -- 300 MHz) and higher (7.8 -- 16.5 GHz) frequencies.  We show that the fading rate has dropped by a factor of $\approx 3$ over the past 50 years at the lower frequencies, while remaining relatively constant at the higher frequencies, which is in agreement with the findings of others.  Our findings at 1405 MHz, in conjunction with a measurement of the fading rate at the nearby frequency of 927 MHz by Vinyajkin (1997), show an intermediate behavior at intermediate frequencies.  We also find that Cassiopeia A, as of $\approx$ 1990, was fading at about the same rate, $\approx 0.6 - 0.7$ \% yr$^{-1}$, at all of these frequencies.  Future measurements are required to determine whether the fading rate will continue to decrease at the lower frequencies, or whether Cassiopeia A will now fade at a relatively constant rate at all of these frequencies.
\end{abstract}


\section{Introduction:  A History of Measurements of Cassiopeia A's Fading Rate}

Cassiopeia A is thought to have exploded around $1681 \pm 15$, based on radial velocity and proper motion measurements of 15 of the supernova remnant's high-velocity [N II]-emitting knots, also called fast-moving flocculi or FMFs (Fesen, Becker, \& Goodrich 1988).  Indeed, Flamsteed (1725) observed a 6th magnitude star within 13$\arcmin$ of Cassiopeia A's location in 1680 (Ashworth 1980) that later became a subject of debate when the star could not be found in the sky (Herschel 1798; Baily 1835).  Cassiopeia A was re-discovered in 1943 at 160 MHz by Reber (1944), although a re-analysis of a day-long strip chart recording published by Jansky (1935) suggests that he unknowingly detected it as early as 1932 at 20.5 MHz (Sullivan 1978).  Accurate measurements of Cassiopeia A's flux density - both absolute measurements and measurements relative to the flux density of the extragalactic radio source Cygnus A - began in 1949 at 81.5 MHz (Ryle \& Elsmore 1951).

\subsection{The Fading Rate of Cassiopeia A around 1965}

H\"ogbom \& Shakeshaft (1961), with Ryle \& Elsmore's (1951) 1949 measurement and two measurements of their own made in 1956 and 1960 also at 81.5 MHz, first showed that Cassiopeia A was fading; they measured a fading rate of $1.06 \pm 0.14$ \% yr$^{-1}$, which was later refined to $1.29 \pm 0.08$ \% yr$^{-1}$ with the addition of four measurements made between 1966 and 1969 also at 81.5 MHz (Scott, Shakeshaft, \& Smith 1969).  
Using data spanning 1957 -- 1972, Baars \& Hartsuijker (1972) measured significantly lower fading rates at 1420 and 3000 MHz, $0.89 \pm 0.12$ and $0.92 \pm 0.15$ \% yr$^{-1}$, respectively; they first suggested that Cassiopeia A was fading at different rates at different frequencies.  By 1977, accurate fading rates had been determined at six different frequencies, spanning 81.5 to 9400 MHz.  Baars et al. (1977; see also Dent, Aller, \& Olsen 1974) collected these measurements, which we re-list in Table 1 and plot in Figure 1, and determined the following empirical equation that describes the frequency dependence of Cassiopeia A's fading rate for the $\approx$ 1965 epoch:
\begin{equation}
-\frac{100}{F_{\nu}}\frac{dF_{\nu}}{dt} = 0.97(\pm0.04)-0.30(\pm0.04)\log{\nu_{GHz}}.
\end{equation}
Using this equation and absolute measurements of Cassiopeia A's flux density made around 1965 over a frequency range that spans 22 MHz to 31 GHz, Baars et al. (1977) also determined an empirical equation that describes the absolute spectrum of Cassiopeia A for the 1965 epoch to an accuracy of $\approx 2$\%.  Empirical equations that describe the absolute spectra of Cygnus A and Taurus A, and a semi-absolute spectrum for Virgo A, were also determined.

\subsection{The Decreasing Fading Rate of Cassiopeia A at Low Radio Frequencies (38 -- 300 MHz)}

Equation (1) was first challenged by Erickson \& Perley (1975) who's
1974 measurement of the flux density of Cassiopeia A at 38 MHz was
brighter than the prediction of Baars et al. (1977) at the 3.5 $\sigma$
level.  Subsequent measurements at 38 MHz by Read (1977a,b), Walczowski
\& Smith (1985), Rees (1990), and Vinyajkin (1997) confirmed that the
fading rate had decreased from $1.9 \pm 0.5$ \% yr$^{-1}$ (determined
using only the 1955 -- 1966 measurements; Read 1977a) to $0.66 \pm
0.17$ \% yr$^{-1}$ (determined using all of the measurements through
1995; Vinyajkin 1997).\footnote{Rees (1990) suggested that the
measurements at 38 MHz, as of 1987, could be treated as being
consistent with a constant fading rate of $\sim 0.8$ \% yr$^{-1}$,
particularly if one ignored the 1949 -- 1969 measurements of Ryle \&
Elsmore (1951), H\"ogbom \& Shakeshaft (1961), and Scott, Shakeshaft,
\& Smith (1969), which imply a significantly higher fading rate, $1.29
\pm 0.08$ \% yr$^{-1}$, at the nearby frequency of 81.5 MHz.  However,
Rees (1990) did not rule out the possibility of a changing fading rate
at 38 MHz, and indeed, the later measurement of this fading rate by
Vinyajkin (1997) suggests that it has changed at the $\approx 2.3$
$\sigma$ confidence level.}

Similar and more accurate results have been found at 81.5 MHz.  As stated above Scott, Shakeshaft, \& Smith (1969) found a 1949 -- 1969 fading rate of $1.29 \pm 0.08$ \% yr$^{-1}$ at this frequency.  Hook, Duffett-Smith, \& Shakeshaft (1992) found a significantly lower fading rate of $0.92 \pm 0.16$ \% yr$^{-1}$ when they included their 1989 measurement, and a fading rate of $\approx 0.63$ \% yr$^{-1}$ when they considered only the measurements made after 1965.  Agafonov (1996) found similar, although slightly higher values:  $1.25 \pm 0.06$ \% yr$^{-1}$ (1949 -- 1985) and $\sim 0.8$ \% yr$^{-1}$ (1973 -- 1985).  

Consistent measurements have also been made at 102 MHz (Agafonov 1996), 151 -- 152 MHz (Read 1977a; Agafonov 1996; Vinyajkin 1997), and 290 -- 300 MHz (Baars \& Hartsuijker 1972; Vinyajkin 1997).  We plot the low frequency (38 -- 300 MHz) measurements of Cassiopeia A's fading rate, as well as the intervals of time over which these measurements were made, in Figure 2, and we list this information in Table 2.  Clearly, as is suggested by Hook, Duffett-Smith, \& Shakeshaft (1992), and advocated by Agafonov (1996), the rate at which Cassiopeia A is fading at low frequencies has decreased over the past 50 years.\footnote{Hook, Duffett-Smith, \& Shakeshaft (1992) suggested that instrumental uncertainties and/or flaring of Cassiopeia A's radio emission might systematically affect measurements of Cassiopeia A's fading rate at these low frequencies.  However, they made arguments to the contrary, and Agafonov (1996) provided evidence to the contrary.  Consequently, we do not further pursue these scenarios in this paper; instead, we refer the interested reader to these papers.}  We find the fading rate to be decreasing by $\approx 2$ \% yr$^{-1}$ per century at these frequencies.  

\subsection{The Constant Fading Rate of Cassiopeia A at High Radio Frequencies (7.8 -- 16.5 GHz)}

However, different conclusions have been reached at significantly higher frequencies.  O'Sullivan \& Green (1999) compare four measurements of Cassiopeia A's relative brightness to Cygnus A that they made in 1994 and 1995 at 13.5, 15.5, and 16.5 GHz to the predictions of Baars et al. (1977).  The measurements of O'Sullivan \& Green (1999) are perfectly consistent with the $\approx 0.6$ \% yr$^{-1}$ predictions of Equation (1),\footnote{Technically, Equation (1) is an extrapolation at frequencies greater than 9.4 GHz; however, this is not a large extrapolation.} suggesting that the rate at which Cassiopeia A fades has not changed significantly over the course of the last half century at these high frequencies.  We plot the high frequency (7.8 -- 16.5 GHz) measurements of Cassiopeia A's fading rate in Figure 3, and we list this information in Table 3.   

These apparently contradictory behaviors at low and high radio
frequencies suggests that the rate at which Cassiopeia A fades is
changing in a frequency-dependent way:  the fading rate is decreasing
at the low frequencies (38 -- 300 MHz) and is relatively constant at
the high frequencies (7.8 -- 16.5 GHz).  In this paper, we investigate
and confirm this trend at the intermediate frequency of 1405 MHz with
measurements of the relative flux density of Cassiopeia A to Cygnus A
that we made between 1995 and 1999 with the 40-foot telescope at the
National Radio Astronomy Observatory in Green Bank, WV.  We analyze
these measurements in \S 2; we draw conclusions in \S 3.

\section{Green Bank 40-foot Telescope Measurements of the Relative Flux Density of Cassiopeia A to Cygnus A at 1405 MHZ from 1995 to 1999}

In August of 1995, 1996, 1997, and 1999, we took drift scans of
Cassiopeia A and Cygnus A using the 40-foot telescope at the National
Radio Astronomy Observatory in Green Bank, WV.  In 1995, we used a 70
MHz band centered about 1405 MHz; in 1996 -- 1999, we used a 110 MHz
band, also centered about 1405 MHz.  The drift scans were $\approx 4$
degrees in length, which is $\approx 4$ times the resolution element of
the telescope.  We describe these observations and their analysis in
greater detail below.  In total, we measured the relative flux density
of Cassiopeia A to Cygnus A, $F_{1405}^{Cas\,A}/F_{1405}^{Cyg\,A}$,
five times over a span of five years.  We list these measurements in
Table 4.  If one models the flux density of Cassiopeia A as a constant
over this interval of time, we find that
$F_{1405}^{Cas\,A}/F_{1405}^{Cyg\,A} = 1.266 \pm 0.023$ for a mean
epoch of 1997.4.  This implies a statistical measurement error of
$\approx 1.8$ \%, which is an upper limit since Cassiopeia A is in
reality fading over this time interval.  Since we used only one of two
orthogonal linears, we add in quadrature to this statistical error,
systematic errors of $\la 0.2$ \% and $\la 0.5$ \%, corresponding to
the polarizations of Cassiopeia A and Cygnus A, respectively.  This
implies a total measurement error of $\la 1.9$ \%.  Conservatively
adopting a total measurement error of 1.9 \%, a value of
$F_{1405}^{Cyg\,A} = 1581$ Jy from the absolute spectrum calibration of
Cygnus A from Baars et al. (1977), and a 2 \% uncertainty in this
calibration, we find that $F_{1405}^{Cas\,A} = 2002 \pm 55$ Jy for this
mean epoch.

Prior to the transit of each source, we manually adjusted the
declination of the telescope, ensuring, with the aid of a strip chart
recorder, that the pointing of the telescope contributed no more than
$\approx 0.5$ \% of error to the flux density measurement.  During
transit and for the remainder of the observation, we left the telescope
at this, its final declination.  Afterward, we excised the data from
the first half of the observation that we had taken at declinations
other than the final declination.  This allowed the background to be
reliably modeled and subtracted.  Across the $\approx 4$ degree lengths
of the observations, the background appears to be linear, both in the
case of the Cassiopeia A observations and in the case of the Cygnus A
observations.  Consequently, we modeled the background as linear and
subtracted it, simultaneously removing any H I component to the
emission.  We estimate that background subtraction contributes $\approx
1$ \% of error to each flux density measurement.  Finally, before and
after the $\approx 30$ minute durations of the Cassiopeia A
observations and the $\approx 20$ minute durations of the Cygnus A
observations, we took two-minute-long calibration readings.  Each pair
of readings agreed to better than $\approx 1$ \%; none-the-less, we
linearly interpolated between the readings, making this a negligible
source of error.  Consequently and in total, we estimate that the
measured flux density ratios, $F_{1405}^{Cas\,A}/F_{1405}^{Cyg\,A}$,
should be in error by no more than $\approx 1.6$ \%; this is consistent
with the statistical measurement error of $\la 1.8$ \% that we
determine above.  Conservatively adopting total measurement errors of
1.9 \%, a 1965 value of $F_{1405}^{Cas\,A}/F_{1405}^{Cyg\,A} = 1.542$
from the absolute spectrum calibrations of Cassiopeia A and Cygnus A
from Baars et al. (1977), and 2 \% uncertainties in each of these
calibrations, we find a 1965 -- 1999 fading rate of $0.62 \pm 0.12$ \%
yr$^{-1}$ ($\chi^2 = 0.36$ for $\nu = 4$ degrees of freedom) at 1405
MHz.  We plot the 1405 MHz light curve in Figure 4.

\section{Conclusions:  A Decreasing Fading Rate of Cassiopeia A at Intermediate Radio Frequencies (927 -- 3060 MHz)}

We plot the intermediate frequency (927 -- 3060 MHz) measurements of Cassiopeia A's fading rate in Figure 5, and we list this information in Table 5.  When compared to measurements of the fading rate between 1957 and 1976, our measurement at 1405 MHz, in conjunction with a recent measurement at the nearby frequency of 927 MHz by Vinyajkin (1997), show that the fading rate is decreasing at a rate that is intermediate to the rates measured at the lower (38 -- 300 MHz) and the higher (7.8 -- 16.5 GHz) frequencies.  We find the fading rate to be decreasing by $\approx 1$ \% yr$^{-1}$ per century at these intermediate frequencies. 
 
Furthermore, we find that around 1990, Cassiopeia A was fading at about the same rate, $\approx 0.6 - 0.7$ \% yr$^{-1}$, at all of these frequencies.  The next decade of observations should reveal whether the fading rate will continue to decrease at the lower frequencies, or whether Cassiopeia A will now fade at a relatively constant rate at all of these frequencies.  

\acknowledgements

This research is supported in part by NASA grant NAG5-2868 and NASA contract NASW-4690.  We are grateful to Rick Fisher and Don Lamb, whose comments greatly improved this paper.  
We are extremely grateful to Sue Ann Heatherly for her dedication to educational and outreach activities at Green Bank, and for making the 40-foot telescope and the facilities at Green Bank available to ERIRA since 1992.  We are also very grateful to Carl Chestnut for the technical assistance he has given us over the years.  We also want to thank Dan Fellows, Randy Bish, and Alan Fuller for their past contributions to ERIRA, and Walter Glogowski and Jeremy Garris for their continued dedication to the program.  Last, but not least, we want to thank the over 100 students and educators who have participated in ERIRA over the past eight years, many of whom helped to collect the data presented in this paper.

\clearpage

\begin{deluxetable}{ccccc}
\tablecolumns{5}
\tablewidth{0pc}
\tablecaption{Fading Rates of Cassiopeia A around 1965}
\tablehead{\colhead{Frequency\tablenotemark{a}} & \colhead{Measurement Epochs} & \colhead{Central Epoch} & \colhead{Fading Rate\tablenotemark{b}} & \colhead{Reference\tablenotemark{c}}}
\startdata
81.5 & 1949 -- 1969 & 1959 & $1.29 \pm 0.08$ & 1 \nl
950 & 1964 -- 1972 & 1968 & $0.85 \pm 0.05$ & 2 \nl
1420 & 1957 -- 1971 & 1964 & $0.89 \pm 0.12$ & 3 \nl
1420 & 1957 -- 1976 & 1966 & $0.86 \pm 0.02$ & 4 \nl
3000 & 1961 -- 1972 & 1966 & $0.92 \pm 0.15$ & 3 \nl
3060 & 1961 -- 1971 & 1966 & $1.04 \pm 0.21$ & 5 \nl
7800 & 1963 -- 1974 & 1968 & $0.70 \pm 0.10$ & 6 \nl
9400 & 1961 -- 1971 & 1966 & $0.63 \pm 0.12$ & 5 \nl
\enddata
\tablenotetext{a}{MHz.}
\tablenotetext{b}{\% yr$^{-1}$.}
\tablenotetext{c}{1. Scott, Shakeshaft, \& Smith 1969; 2. Stankevich, Ivanov, \& Torkhov 1973; 3. Baars \& Hartsuijker 1972; 4. Read 1977a; 5. Stankevich et al. 1973; 6. Dent, Aller, \& Olsen 1974.}
\end{deluxetable}

\clearpage

\begin{deluxetable}{ccccc}
\tablecolumns{5}
\tablewidth{0pc}
\tablecaption{Fading Rates of Cassiopeia A at Low Radio Frequencies (38 -- 300 MHz)}
\tablehead{\colhead{Frequency\tablenotemark{a}} & \colhead{Measurement Epochs} & \colhead{Central Epoch} & \colhead{Fading Rate\tablenotemark{b}} & \colhead{Reference\tablenotemark{c}}}
\startdata
38 & 1955 -- 1966 & 1960 & $1.9 \pm 0.5$ & 1 \nl
38 & 1955 -- 1995 & 1975 & $0.66 \pm 0.17$ & 2 \nl
81.5 & 1949 -- 1960 & 1954 & $1.06 \pm 0.14$ & 3 \nl
81.5 & 1949 -- 1969 & 1959 & $1.29 \pm 0.08$ & 4 \nl
81.5 & 1949 -- 1985 & 1967 & $1.25 \pm 0.06$ & 5 \nl
81.5 & 1949 -- 1989 & 1969 & $0.92 \pm 0.16$ & 6 \nl
81.5 & 1966 -- 1989 & 1978 & $\approx 0.63\tablenotemark{d}$ & 6 \nl
102 & 1977 -- 1993 & 1985 & $0.80 \pm 0.12$ & 5 \nl
151 & 1966 -- 1976 & 1971 & $1.2 \pm 0.4$ & 1 \nl
152 & 1966 -- 1985 & 1975 & $1.06 \pm 0.15$ & 5 \nl
151.5 & 1966 -- 1994 & 1980 & $0.86 \pm 0.09$ & 2 \nl
151.5 & 1980 -- 1994 & 1988 & $1.11 \pm 0.22$ & 2 \nl
300 & 1961 -- 1971 & 1966 & $0.91 \pm 0.15$ & 7 \nl
290 & 1978 -- 1996 & 1988 & $0.66 \pm 0.07$ & 2 \nl
\enddata
\tablenotetext{a}{MHz.}
\tablenotetext{b}{\% yr$^{-1}$.}
\tablenotetext{c}{1. Read 1977a; 2. Vinyajkin 1997; 3. H\"ogbom \& Shakeshaft 1961; 4. Scott, Shakeshaft, \& Smith 1969; 5. Agafonov 1996; 6. Hook, Duffett-Smith, \& Shakeshaft 1992; 7. Baars \& Hartsuijker 1972.}
\tablenotetext{d}{We calculate an error of $\pm 0.17$ from information provided in their paper and in Scott, Shakeshaft, \& Smith 1969.}
\end{deluxetable}

\clearpage

\begin{deluxetable}{ccccc}
\tablecolumns{5}
\tablewidth{0pc}
\tablecaption{Fading Rates of Cassiopeia A at High Radio Frequencies (7.8 -- 16.5 GHz)}
\tablehead{\colhead{Frequency\tablenotemark{a}} & \colhead{Measurement Epochs} & \colhead{Central Epoch} & \colhead{Fading Rate\tablenotemark{b}} & \colhead{Reference\tablenotemark{c}}}
\startdata
7.8 & 1963 -- 1974 & 1968 & $0.70 \pm 0.10$ & 1 \nl
9.4 & 1961 -- 1971 & 1966 & $0.63 \pm 0.12$ & 2 \nl
13.5 & 1965 -- 1994 & 1980 & $\approx 0.63\tablenotemark{d}$ & 3 \nl
15.5 & 1965 -- 1994 & 1980 & $\approx 0.61\tablenotemark{d}$ & 3 \nl
16.5 & 1965 -- 1995 & 1980 & $\approx 0.60\tablenotemark{d}$ & 3 \nl
\enddata
\tablenotetext{a}{GHz.}
\tablenotetext{b}{\% yr$^{-1}$.}
\tablenotetext{c}{1. Dent, Aller, \& Olsen 1974; 2. Stankevich et al. 1973; 3. O'Sullivan \& Green 1999.}
\tablenotetext{d}{We estimate errors of $\pm 0.14$ from information provided in their paper.}
\end{deluxetable}

\clearpage

\begin{deluxetable}{ccc}
\tablecolumns{3}
\tablewidth{0pc}
\tablecaption{Green Bank 40-foot Telescope Measurements of the Relative Flux Density of Cassiopeia A to Cygnus A at 1405 MHZ from 1995 to 1999}
\tablehead{\colhead{Epoch} & \colhead{$F_{1405}^{Cas\,A}/F_{1405}^{Cyg\,A}$\tablenotemark{a}} & \colhead{$F_{1405}^{Cas\,A}$\tablenotemark{b}}}
\startdata
1995.6 & 1.294 & $2046 \pm 56$ \nl
1996.6 & 1.267 & $2003 \pm 55$ \nl
1997.6 & 1.273 & $2013 \pm 56$ \nl
1997.6 & 1.265 & $2000 \pm 55$ \nl
1999.6 & 1.231 & $1946 \pm 54$ \nl
\enddata
\tablenotetext{a}{$\pm 1.9$ \% (see \S 2).}
\tablenotetext{b}{Jy; for $F_{1405}^{Cyg\,A} = 1581 \pm 32$ Jy (see \S 2).}
\end{deluxetable}

\clearpage

\begin{deluxetable}{ccccc}
\tablecolumns{5}
\tablewidth{0pc}
\tablecaption{Fading Rates of Cassiopeia A at Intermediate Radio Frequencies (927 -- 3060 MHz)}
\tablehead{\colhead{Frequency\tablenotemark{a}} & \colhead{Measurement Epochs} & \colhead{Central Epoch} & \colhead{Fading Rate\tablenotemark{b}} & \colhead{Reference\tablenotemark{c}}}
\startdata
950 & 1964 -- 1972 & 1968 & $0.85 \pm 0.05$ & 1 \nl
927 & 1977 -- 1996 & 1986 & $0.73 \pm 0.05$ & 2 \nl
1420 & 1957 -- 1971 & 1964 & $0.89 \pm 0.12$ & 3 \nl
1420 & 1957 -- 1976 & 1966 & $0.86 \pm 0.02$ & 4 \nl
1405 & 1965 -- 1999 & 1982 & $0.62 \pm 0.12$ & 5 \nl
3000 & 1961 -- 1972 & 1966 & $0.92 \pm 0.15$ & 3 \nl
3060 & 1961 -- 1971 & 1966 & $1.04 \pm 0.21$ & 6 \nl
\enddata
\tablenotetext{a}{MHz.}
\tablenotetext{b}{\% yr$^{-1}$.}
\tablenotetext{c}{1. Stankevich, Ivanov, \& Torkhov 1973; 2. Vinyajkin 1997; 3. Baars \& Hartsuijker 1972; 4. Read 1977a; 5. this paper; 6. Stankevich et al. 1973.}
\end{deluxetable}

\clearpage

\figcaption[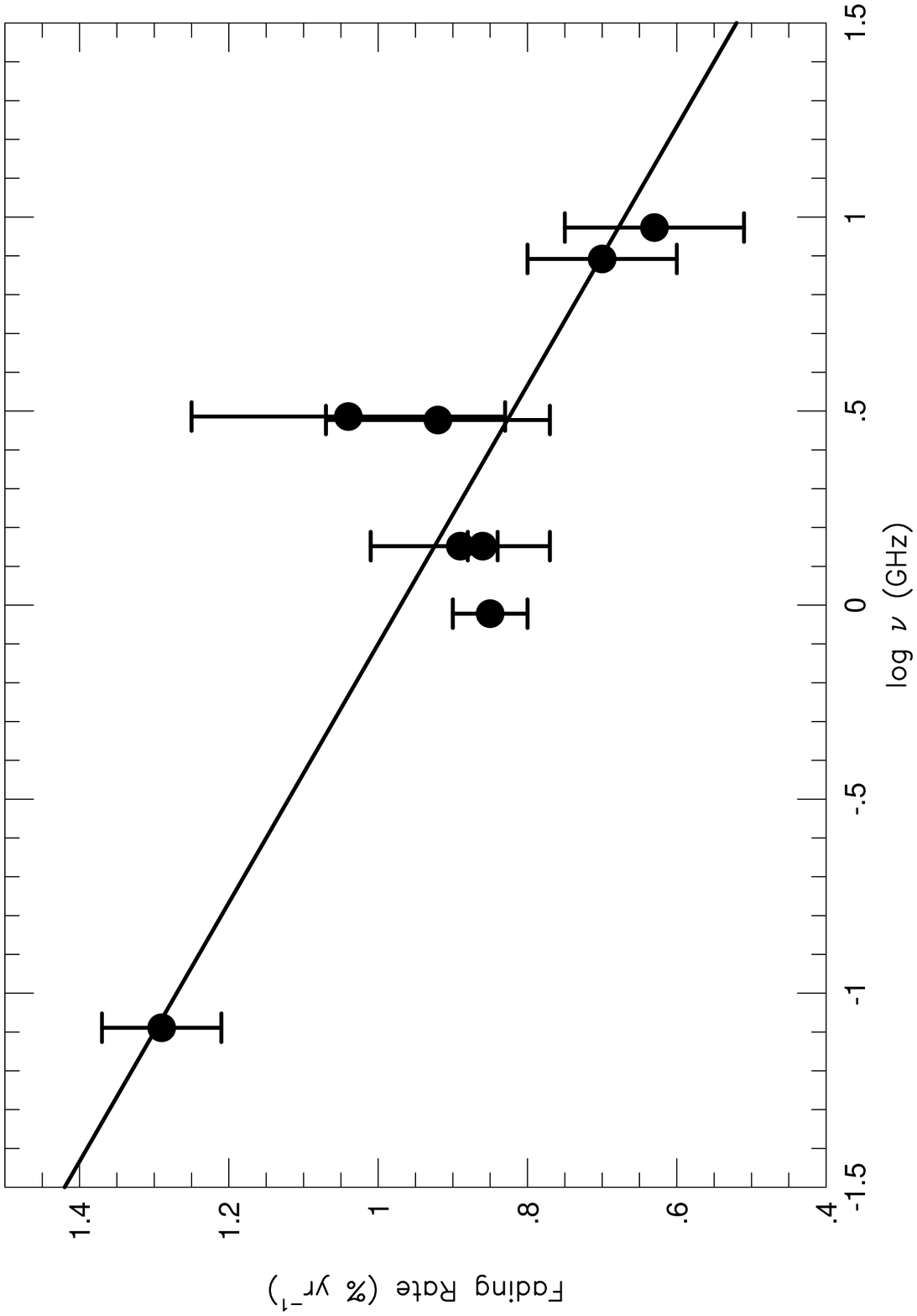]{Fading rates of Cassiopeia A as a function of frequency for the $\approx 1965$ epoch, from Baars et al. (1977; see Table 1).  Equation (1), also from Baars et al. (1977), is plotted as the solid line.\label{cas1.ps}}

\figcaption[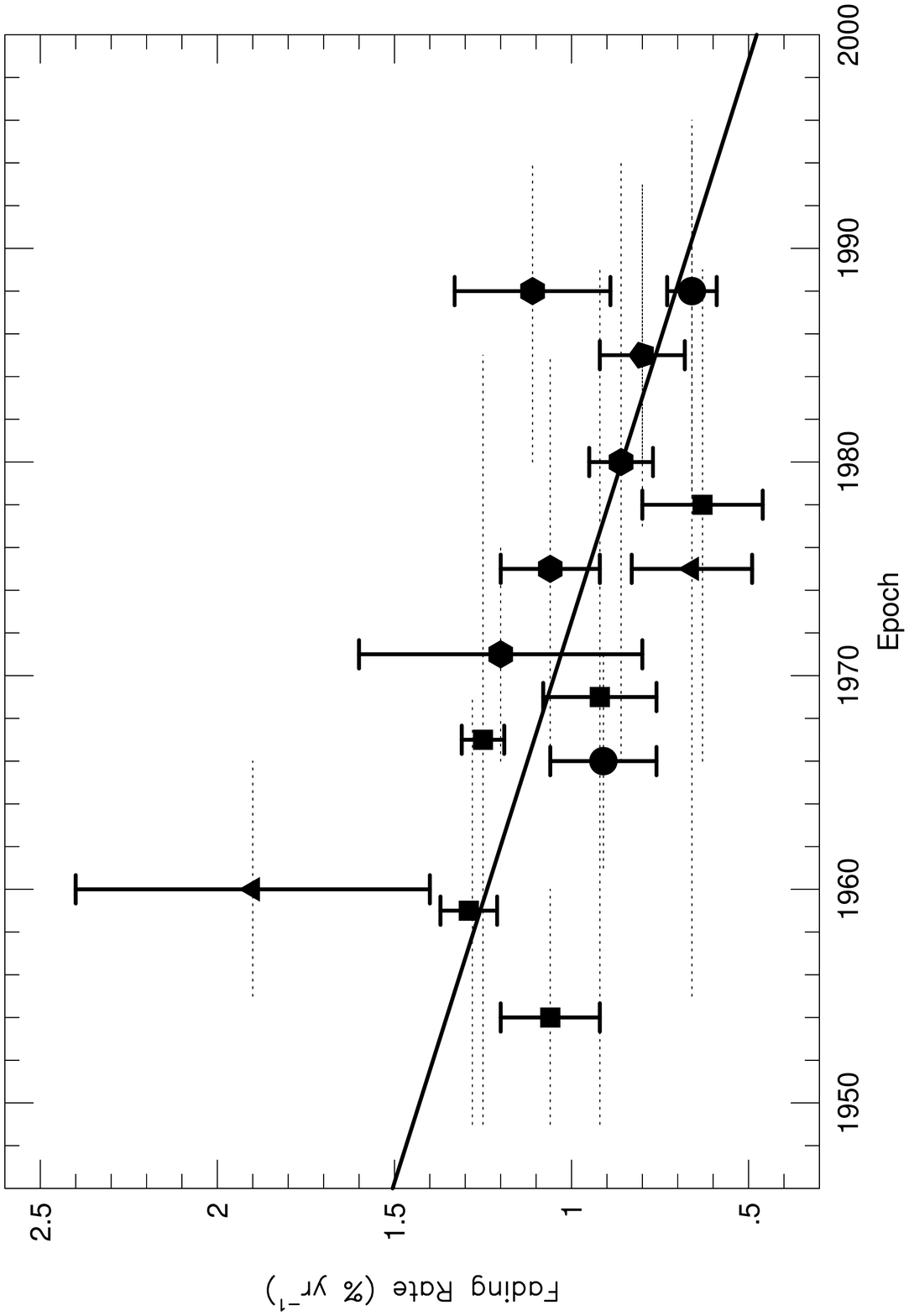]{Fading rates of Cassiopeia A as a function of epoch for the low radio frequencies (38 -- 300 MHz; see Table 2).  The dotted lines indicate over what interval of time measurements were taken to determine each fading rate.  Triangles denote 38 MHz measurements, squares denote 81.5 MHz measurements, pentagons denote 102 MHz measurements, hexagons denote 151 -- 152 MHz measurements, and circles denote 290 -- 300 MHz measurements.  The solid line corresponds to a fading rate that is decreasing at a rate of $\approx 2$ \% yr$^{-1}$ per century.\label{cas2.ps}}

\figcaption[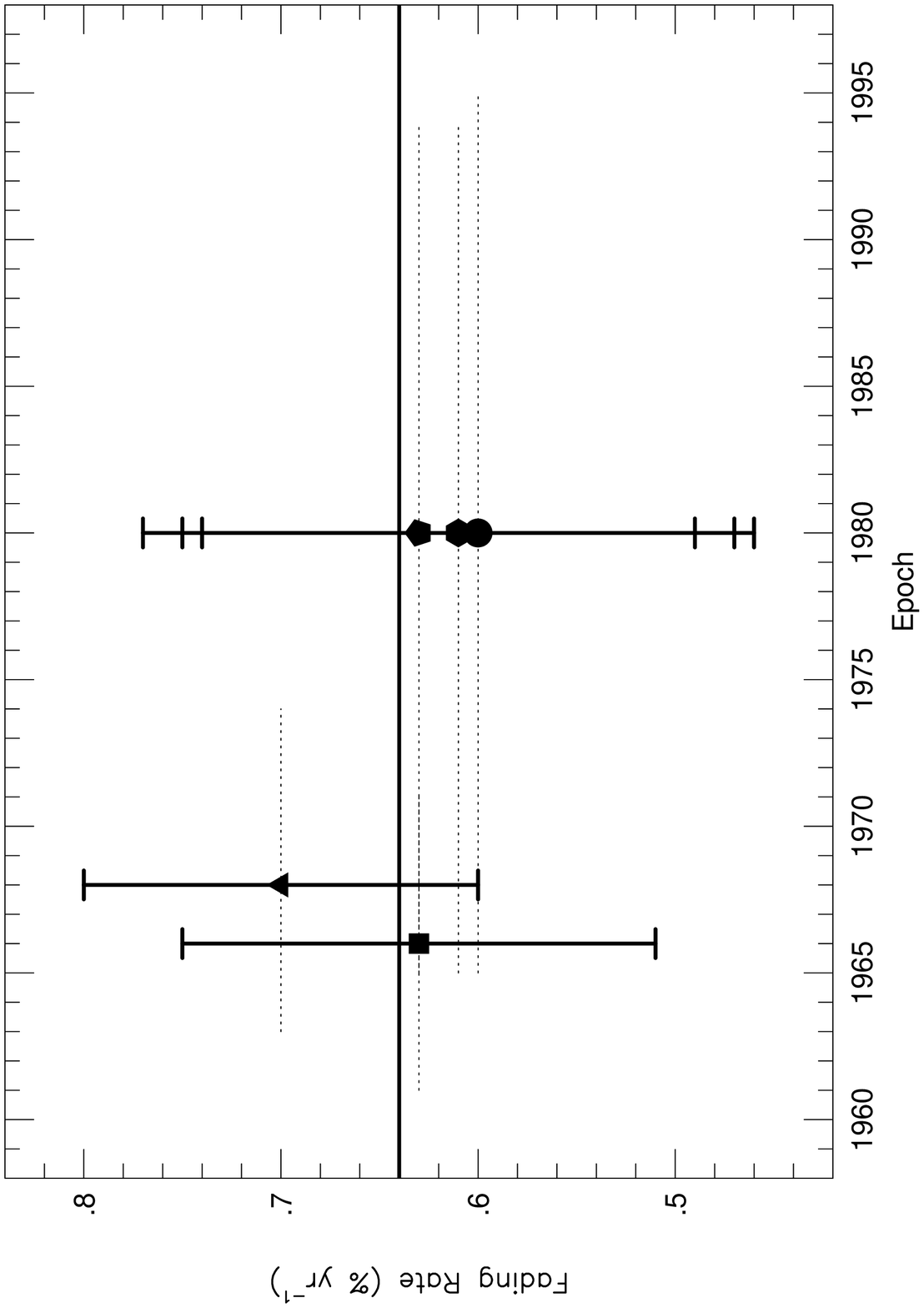]{Fading rates of Cassiopeia A as a function of epoch for the high radio frequencies (7.8 -- 16.5 GHz; see Table 3).  The dotted lines are as defined in Figure 2.  Triangles denote 7.8 GHz measurements, squares denote 9.4 GHz measurements, pentagons denote 13.5 GHz measurements, hexagons denote 15.5 GHz measurements, and circles denote 16.5 GHz measurements.  The solid line corresponds to a constant fading rate.\label{cas3.ps}}

\figcaption[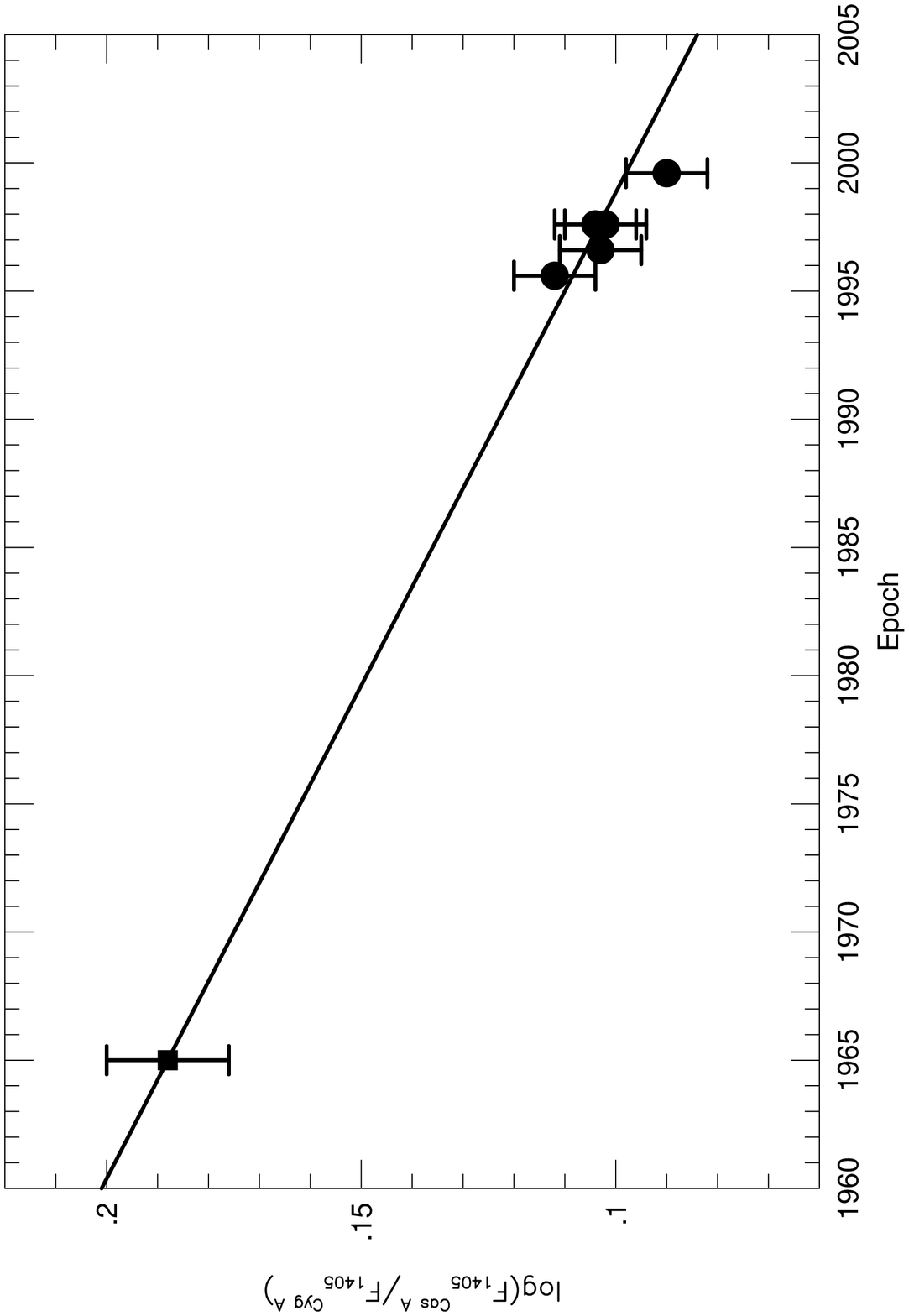]{Light curve of measurements of the relative flux density of Cassiopeia A to Cygnus A at 1405 MHz.  Circles denote our measurements (see Table 4); the square denotes the 1965 value from Baars et al. (1977).  The solid line corresponds to a fading rate of 0.62 \% yr$^{-1}$.\label{cas4.ps}}

\figcaption[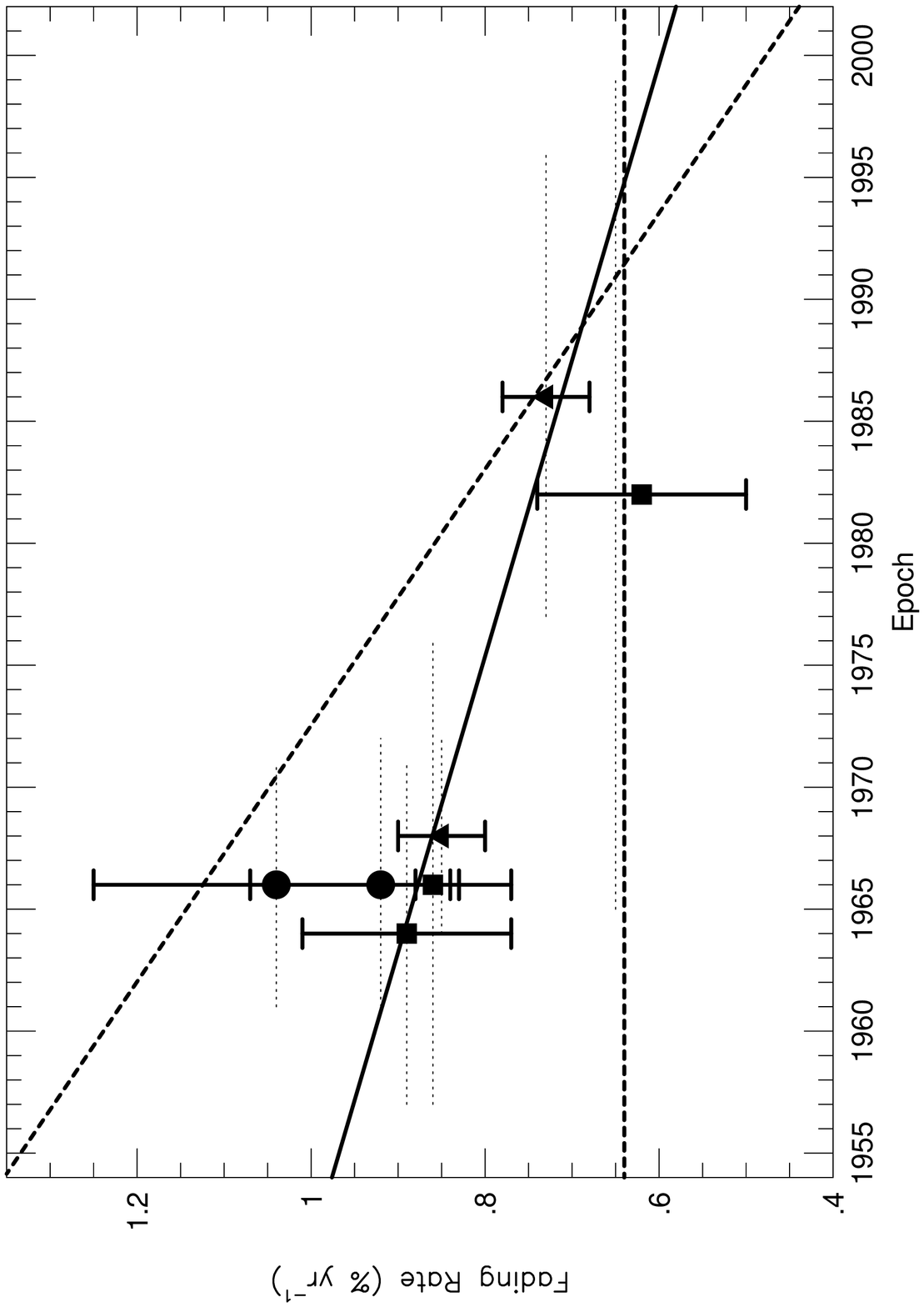]{Fading rates of Cassiopeia A as a function of epoch for the intermediate radio frequencies (927 -- 3060 MHz; see Table 5).  The dotted lines are as defined in Figure 2.  Triangles denote 927 -- 950 MHz measurements, squares denote 1405 -- 1420 MHz measurements, and circles denote 3000 -- 3060 MHz measurements.  The solid line corresponds to a fading rate that is decreasing at a rate of $\approx 1$ \% yr$^{-1}$ per century.  The dashed lines are the same as the solid lines in Figures 2 and 3.\label{cas5.ps}}

\clearpage

\setcounter{figure}{0}

\begin{figure}[tb]
\plotone{cas1.ps}
\end{figure}

\begin{figure}[tb]
\plotone{cas2.ps}
\end{figure}

\begin{figure}[tb]
\plotone{cas3.ps}
\end{figure}

\begin{figure}[tb]
\plotone{cas4.ps}
\end{figure}

\begin{figure}[tb]
\plotone{cas5.ps}
\end{figure}


\begin{thebibliography}{}
\bibitem[Herschel 1798]{h98}
Herschel, C. 1798, Catalog of stars, taken from Flamsteed's observations...and not inserted in the British Catalog. With an index [and] a collection of errata (London)
\end{thebibliography}
\end{document}